\documentclass[sn-mathphys-num]{sn-jnl}
\usepackage{graphicx}%
\usepackage{subcaption}%
\usepackage{multirow}%
\usepackage{amsmath,amssymb,amsfonts}%
\usepackage{amsthm}%
\usepackage{mathrsfs}%
\usepackage[title]{appendix}%
\usepackage{xcolor}%
\usepackage{color}%
\usepackage{soul}%
\usepackage{textcomp}%
\usepackage{manyfoot}%
\usepackage{booktabs}%
\usepackage{algorithm}%
\usepackage{algorithmicx}%
\usepackage{algpseudocode}%
\usepackage{listings}%
\usepackage{orcidlink}
\usepackage{tikz}
\usetikzlibrary{backgrounds}

\begin{document}

\title[Measurements of power dissipated in an atmospheric pressure plasma jet device with double plasma discharge ignition]{Measurements of power dissipated in an atmospheric pressure plasma jet device with double plasma discharge ignition}

\author*[1]{\fnm{Fellype} \spfx{do} \sur{Nascimento}\orcidlink{0000-0002-8641-9894}}

\author[1]{\fnm{Kleber A.} \sur{Petroski}\orcidlink{0009-0009-7055-6841}}

\author[2]{\fnm{Thalita M.} \sur{C. Nishime}\orcidlink{0000-0003-2844-3156}}

\author[1]{\fnm{Konstantin G.} \sur{Kostov}\orcidlink{0000-0002-9821-8088}}

\affil*[1]{\orgdiv{Faculty of Engineering and Sciences}, \orgname{São Paulo State University–UNESP}, \orgaddress{\city{Guaratinguetá}, \postcode{12516-410}, \country{Brazil}}}

\affil[2]{\orgname{Leibniz Institute for Plasma Science and Technology–INP}, \orgaddress{\city{Greifswald}, \postcode{17489}, \country{Germany}}}

\abstract{Atmospheric pressure plasma jets (APPJs) are versatile devices with numerous applications. This work focuses on APPJs generated at the tip of long, flexible tubes using the jet transfer technique. The plasma source consists of a primary discharge and a secondary discharge forming the plasma jet. Discharge power measurements were carried out in a way that it was possible to separate the contribution of the primary discharge from the total power dissipated by the plasma source. Both power and effective current were analyzed under different operating conditions. The results show that the variation of the primary discharge power is much lower than the power dissipated by the plasma jet. Additionally, the electrical characteristics of the plasma device were analyzed. Notable differences were observed between the negative and positive phases of the discharge, with a more resistive load in the negative one, which suggests that the electrical equivalent circuit model changes according to the voltage polarity.}




\maketitle

\tikzstyle{background rectangle}=[thin,draw=black]
\begin{tikzpicture}[show background rectangle]

\node[align=justify, text width=0.9*\textwidth, inner sep=1em]{
{This document corresponds to the accepted version of the paper published in \textit{The European Physical Journal D}. If you find this preprint useful for your research and want to cite it in your work, please, refer to the published version:
\\
\\F. do Nascimento, K. A. Petroski, T. M. C. Nishime and K. G. Kostov, \textit{The European Physical Journal D}, Vol. 78 (12), p. 154, (2024) -- DOI: \href{https://dx.doi.org/10.1140/epjd/s10053-024-00946-z}{10.1140/epjd/s10053-024-00946-z}.
\\
 \\You can also find the full text of the published version at: \href{https://rdcu.be/d4Emg}{https://rdcu.be/d4Emg}}
};

\node[xshift=3ex, yshift=-0.7ex, overlay, fill=white, draw=white, above
right] at (current bounding box.north west) {
\textit{Dear reader,}
};

\end{tikzpicture}

\section{Introduction}\label{secIntroduction}

The development of plasma sources for generation of atmospheric pressure plasmas (APPs) remains a highly active area of research, especially for the generation of low temperature plasmas (LTPs) suitable for applications in plasma medicine, surface sterilization/decontamination, agriculture, reduction of chemicals and pharmaceuticals in sewage and treated water, etc. \cite{bartis_interaction_2016, Cvelbar_white_2019, Laroussi_cold_2020, Miebach_medical_2022, adesina_review_2024}. Applications in processing and synthesis of materials also employ APPs at least as part of the procedures \cite{dato_graphene_2019, Nimbekar_plasma_2022, Dufour_basics_2023, Hou_recent_2023}. LTPs can be generated in both open and close environments through different processes and at diverse work pressures, and can form large volume discharges or plasma jets \cite{kim_comparative_1995, liu_review_2022}. Regarding the use of plasma jets, the atmospheric pressure plasma jets (APPJs) have been used successfully in several areas due to their versatility and ease of application \cite{Cvelbar_white_2019, Corbella_flexible_2023}. Particularly, the APPJs formed at the end tip of long tubes can be easily directed to targets that are far from the voltage source. In this way, the application safety is increased and the plasma jets can be applied in difficult-to-access places, such as cavities and the interior of the human body \cite{Corbella_flexible_2023}.

Plasma sources that employ long and flexible tubes to generate a plasma jet at tube tip can be assembled in different ways. One of the very first works reporting the generation of a plasma jet at the tip of a long tube was Hong \textit{et al} \cite{Hong_long_2007}. In that work, the authors just attached the long tube to a needle connected to a power supply and to a gas inlet line. Thus, the applied voltage was increased until a plasma column of 60 cm was formed and the plasma jet was extracted from the end tip of the long tube. Another way to produce APPJs at the tip of long tubes is to make an assembly in a way that the high voltage (HV) electrode ends close to the plasma outlet \cite{Polak_innovative_2012, Wang_propagation_2015, Winter_development_2018, Kurosawa_endoscopic_2019}. In this case the length of the set formed by the plastic tube and the HV electrode is limitless, in theory. However, this technique has as a disadvantage the presence of a HV electrode close to the target to be treated. Thus, the voltage amplitude must be limited and good electrical insulation needs to be done in order to prevent undesirable high-current discharges. Current limiters could also be employed in this case. Another possible way to generate APPJs in the tip of flexible tubes is by using the jet transfer technique \cite{kostov_transfer_2015, Xia_transfer_2016, Bastin_optical_2020, do_nascimento_different_2023}. In this last method, the plasma source is composed by a reactor, usually using a dielectric barrier discharge (DBD) configuration, to which the long tube is attached. This kind of device works in a way that two plasma discharges are ignited using the same working gas, which results in two sources of power dissipation.

A better understanding of the functioning principles of APPJ devices is essential for the development of more efficient and safer plasma sources. The electrical mechanisms employed in the plasma sources are the most important part of those, since APPJ are ignited only if appropriate voltage signals are applied. In this way, electrical characterization of plasma sources is crucial. The electrical characterization of a plasma source can be divided into obtaining discharge parameters and device features, being that the main electrical discharge parameters are the power dissipated in the plasma and the current flowing through the system. Regarding the device features, each kind of design can lead to distinct intrinsic features like the system and the dielectric capacitances and system impedance, which, in turn, affect the electrical equivalent circuit (EEC) model of the plasma source. Many efforts have been made over time to model the electrical equivalent circuit of plasma sources. Although the modeling of parallel plate DBD systems is well established, the modeling of plasma jets is still under study, with different studies reporting, in some cases, very different models \cite{Valdivia-Barrientos_analysis_2006, Monge-Dauge_experimental_2012, Fang_electrical_2012, Pipa_simplest_2012, Jiang_experimental_2013, Eid_experimental_2013, Fang_discharge_2016, Wang_investigation_2018, Pipa_equivalent_2019, Bastin_electrical_2023, Cui_Modified_2023}. In a recent study conducted by Bastin \textit{et al}, modeling of the EEC of a long, flexible tube plasma jet that employs the jet transfer technique was reported \cite{Bastin_electrical_2023}. In that work, the Lissajous curves formed by the applied voltage ($V(t)$) and measured charge ($q(t)$) (also known as $q-V$ plots) from experimental data and simulation using the proposed EEC model presented a good agreement, which indicates that the EEC model is very close to the real experimental set.

In this work, a plasma source that uses the jet transfer technique to ignite a plasma jet at the distal end of a long and flexible tube was characterized in order to obtain the power dissipated on each of the two ignited discharges. For this purpose, power measurements were carried out with the plasma plume on and off. In the first case, the total discharge power was obtained, while with the plasma plume off only the power of the primary discharge was measured. By using this method, it was found that in most cases the power dissipated in the primary discharge changes much less than the power dissipated on the plasma jet. Analysis of the $q-V$ plots and estimations of dielectric and system capacitance were also carried out in this work. From the measured $q-V$ plots it was found that the discharge formed with the plasma plume on presents a highly resistive load when both V and q have negative values, which indicates that the model proposed in \cite{Bastin_electrical_2023} does not apply to the configuration used in this work, in the mentioned condition, but possibly a transmission line model reported in a previous work of that research team \cite{Bastin_optical_2020} does.

\section{Background theory}\label{secTheory}
\begin{figure}
  \centering
  \includegraphics[width=\linewidth]{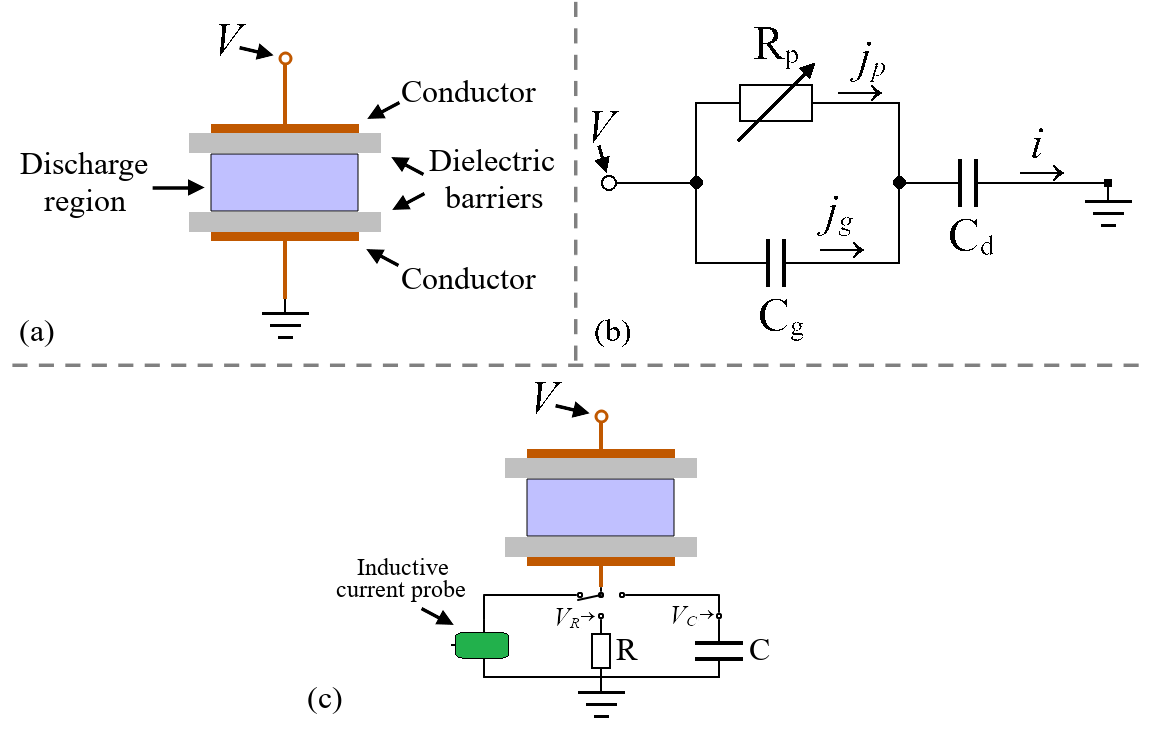}
  \caption{Simple DBD model. (a) Parallel plate DBD device. (b) Corresponding electrical equivalent circuit. (c) Possible setups for measurements of discharge current and power.}
  \label{simpleDBD}
\end{figure}

One of the biggest challenges in describing electrical properties of APP is that plasmas are nonlinear conducting media. Because of that, equivalent circuit analysis of plasma discharges are always simplified and the plasmas are described as a variable resistor. Figure~\ref{simpleDBD} is a parallel plate dielectric barrier discharge (DBD) device and a simplest equivalent circuit used to describe it \cite{Pipa_simplest_2012}. In that circuit, the plasma discharge is represented as a resistor with variable resistance ($R_p$). $C_g$ and $C_d$ represent the capacitance of the gas gap and dielectric material, respectively, and the series association of both capacitors leads to an equivalent capacitance that is commonly referred as reactor cell capacitance or system capacitance ($C_s$), with:

\begin{equation}
  {C_s = \frac{C_g C_d}{C_g + C_d}}\label{eq01}
\end{equation}

The equivalent circuit in Fig.~\ref{simpleDBD} is considered a good approach for the parallel plate DBD device when the applied voltage ($V$) is sinusoidal. When applying $V$ to the circuit and the discharge ignites, the total current ($i$) is the sum of the one passing through the plasma ($j_p$) and the current through the gas gap capacitance ($j_g$), that is:

\begin{equation}
 {i(t) = j_p(t) + j_g(t)}\label{eq02}
\end{equation}

The waveform of $i(t)$ is usually measured with an inductive current probe or by using the shunt resistor method and measuring the voltage signal across the resistor ($V_R (t)$), being that in any case the probe or the shunt resistor is placed close to the ground node, like in Fig.~\ref{simpleDBD}(c) \cite{Ashpis_progress_2017}. It is also possible to measure $i(t)$ by employing the monitor capacitor method and measuring the voltage ($V_C (t)$) at the node. In the last case, $i(t)$ is calculated as:

\begin{equation}
 {i(t) = C \frac{dV_C (t)}{dt} = \frac{dq(t)}{dt}}\label{eq03}
\end{equation}

\noindent where $C$ is the capacitance value of the capacitor.

For the DBD device shown in Fig.~\ref{simpleDBD}, the power dissipated in the discharge can be obtained from the waveforms of applied voltage and total current by applying \cite{Ashpis_progress_2017, Pipa_simplest_2012, Pipa_equivalent_2019}:

\begin{equation}
 {P_{dis} = \frac{1}{\tau} E_{dis} = \frac{1}{\tau} \int_{t_0}^{t_0 + \tau} V(t) i(t) dt }\label{eq04}
\end{equation}

\noindent where $E_{dis}$ is the energy dissipated by the plasma discharge over a period $\tau$. Alternatively, the $q(t)$ waveform can be used instead of $i(t)$. In this case, equation \ref{eq04} turns:

\begin{equation}
{P_{dis} = \dfrac{1}{\tau} \oint q dV}\label{eq05}
\end{equation}

\noindent That is, $E_{dis}$ is the area enclosed in the curve formed by a $q-V$ plot over a period $\tau$. Since both voltage and charge are parametric signals as a function of $t$, for practical purposes, the Green’s theorem can be applied to \ref{eq05} and $P_{dis}$ can be written as:

\begin{equation}
{P_{dis} = \dfrac{1}{2 \tau} \int_{t_0}^{t_0 + \tau} \left[ V(t) q^{\prime}(t) - V^{\prime}(t) q(t) \right] dt}\label{eq06}
\end{equation}

\noindent
where $V^{\prime} (t)$ and $q^{\prime} (t)$ are the time derivatives of $V(t)$ and $q(t)$, respectively. Equation \ref{eq06} has the advantage of being more simple to be used for numerical integration of the discrete voltage and charge signals. It is important to mention that this method for determining $P_{dis}$ following the approach outlined in refs. \cite{Pipa_simplest_2012}, \cite{Pipa_equivalent_2019},and \cite{Ashpis_progress_2017} assumes that the input and output currents are identical. In this way, it disregards other forms of energy dissipation like dielectric losses or electromagnetic radiation.

\begin{figure}
\centering
\begin{subfigure}{0.46\textwidth}
\centering
\includegraphics[width=\textwidth]{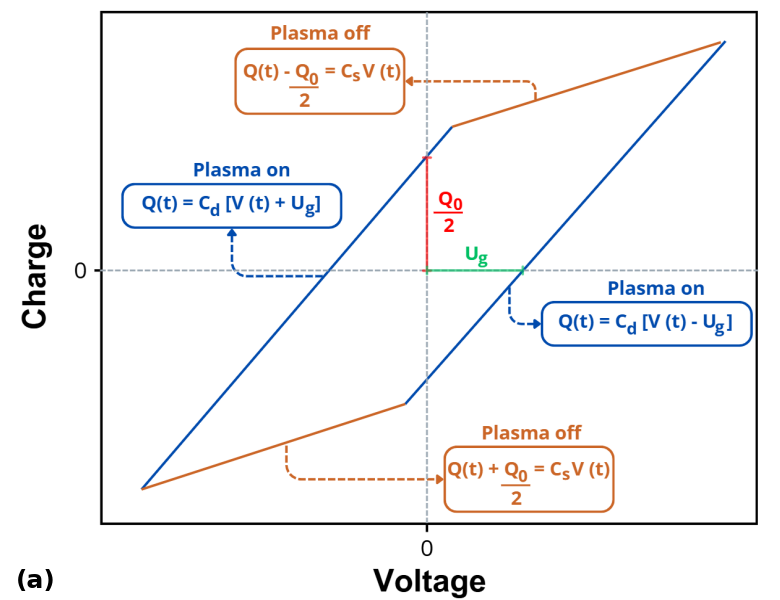}
\end{subfigure}
\begin{subfigure}{0.46\textwidth}
\centering
\includegraphics[width=\textwidth]{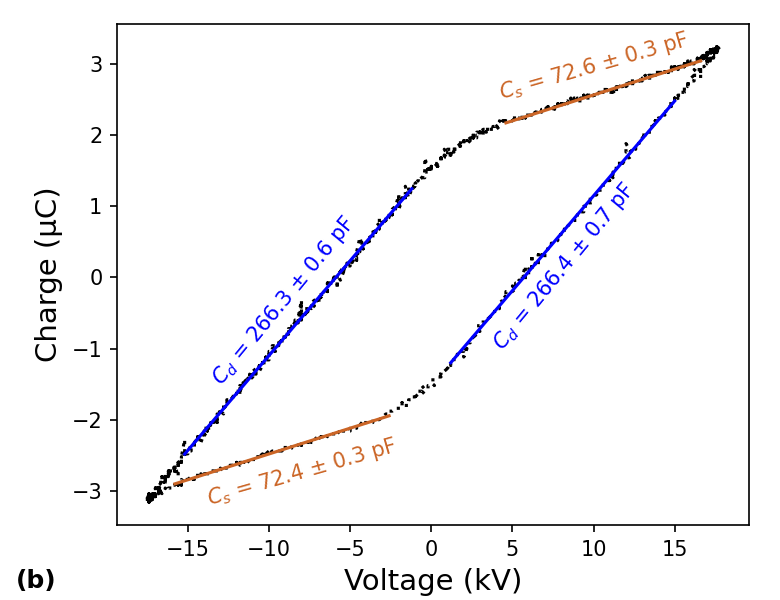}
\end{subfigure}
\caption{Typical $q-V$ plots obtained for parallel plate DBD devices whose discharges are generated using sinusoidal voltage waveform: (a) ideal curve with the equations that describes it in the classical theory and (b) example of a real measurement for the device presented in \cite{Kostov_study_2013}.}
\label{qVexamples}
\end{figure}

Besides the discharge power, the $q-V$ plots can also be used to obtain the values of $C_s$ and $C_d$.
$C_s$ can be extracted from the inclination of the $q-V$ curve in the plasma off phases while $C_d$ is linked to the inclination of the plasma on phases \cite{Pipa_equivalent_2019}, as it is depicted in Fig.~\ref{qVexamples} (a), with real $C_s$ and $C_d$ measurements presented in Fig.~\ref{qVexamples} (b).

Consider for instance the DBD configuration shown in the Fig.~\ref{simpleDBD} powered with a sinusoidal voltage of sufficient amplitude to cause the discharge breakdown. If the electrical discharge formed covers all the interfaces between the dielectrics and the plasma in the region between the electrodes, and Ug is constant for the entire period that the plasma is on. The variation $q(t)$ with $V(t)$ during a complete sinusoidal cycle of $V$ forms the parallelogram shown in figure Fig.~\ref{qVexamples} (a). The relation between $q$ and $V$ of the system runs counterclockwise around the perimeter of this parallelogram, passing through the four straight lines that form it. These straight lines correspond to the moments with and without plasma in the system during a complete sinusoidal cycle of $V$. In the two periods when there is no plasma in the system, the slope of the straight lines (in dark yellow) formed by $q-V$ are equal to $C_s$. And at the two periods when there is no plasma, the slope of the straight lines (in blue) formed by $q-V$ is equal to $C_d$. The same relationship can also be obtained using the EEC shown in Figure~\ref{simpleDBD} (b). However, if the discharge does not fully involve the interfaces between the dielectric and the volume of gas between the electrodes, or if there are parasitic capacitances, the slopes of the straight lines that form the parallelogram when plotting $q-V$ will not be equal to $C_s$ and $C_d$, but will still be related to them.

Although the calculation of discharge power and capacitance values using $q-V$ plots were initially defined for sinusoidal voltage waveforms, those concepts can be applied to any kind of voltage waveform used to ignite a plasma in a DBD device \cite{Pipa_equivalent_2019}. However, it will not always be possible to obtain the values of $C_s$ and/or $C_d$ by this method because those calculations require linear $q-V$ segments in, at least, part of the plasma off and plasma on phases, which may not be possible if the discharge presents a highly resistive load \cite{Kogelschatz_dielectric-barrier_2003}. In addition, the obtention of $C_d$ values from the slope of the $q-V$ curve in the plasma on phase is possible only for $q-V$ plots that resemble a parallelogram. However, if the $q-V$ plot has any other shape $C_d$ needs to be obtained in another way. An alternative method to obtain such a value using $q-V$ curves that differ from a parallelogram was proposed by Pipa \textit{et al} \cite{Pipa_role_2013}. Such a method considers the data points at which the value of charge in the system is maximum and is followed by the extinction of the discharge. At this moment it can be written \cite{Pipa_role_2013}:

\begin{equation}
 {q_{max} = C_d (V_{q_{max}} - U_{res})}\label{eq07}
\end{equation}

\noindent
where $q_{max}$ is the maximum charge value in the cycle, $V_{q_{max}}$ is its associated voltage value and $U_{res}$ is the residual voltage when the discharge is about to extinguish. Thus, if $U_{res}$ is independent of the amplitude of the applied voltage, the $C_d$ value can be extracted from the slope of the straight line formed by the $q_{max}$ and $V_{q_{max}}$ data. A limitation of this method is that it requires the obtention of $q-V$ plots measured with different amplitudes of the applied voltage.

\section{Materials and methods}\label{secMM}

\begin{figure}
\centering
\includegraphics[width=\linewidth]{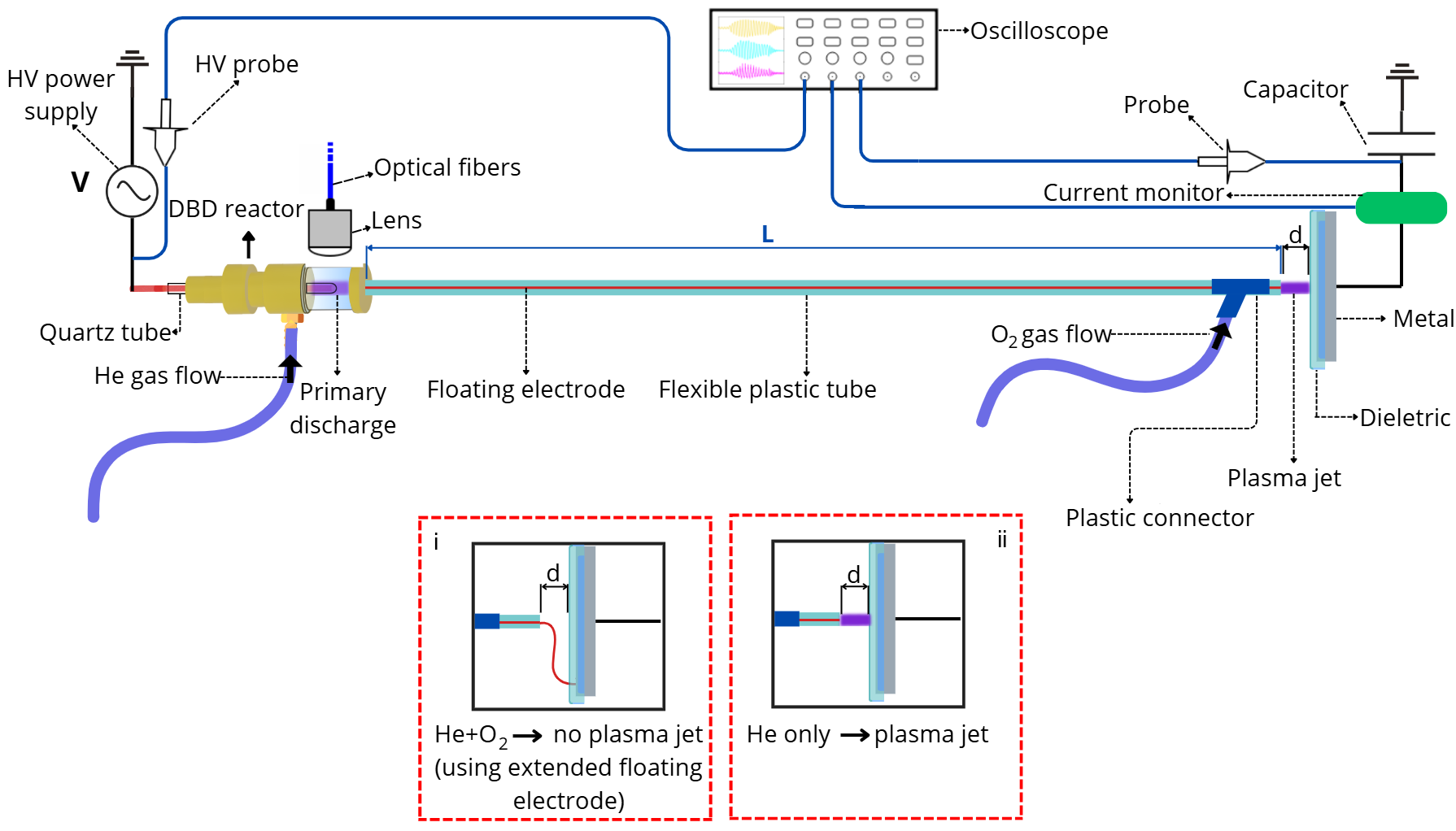}
\caption{Overview of the experimental setup (above) and details for the conditions with (i) plasma jet off and (ii) plasma jet on.}
\label{expSetup}
\end{figure}

An overview of the experimental setup used in this work is presented in Fig.~\ref{expSetup}. The entire plasma source is composed of a DBD reactor to which is attached a long plastic tube (made of low density polyethylene) with a thin (0.5 mm in diameter) and flexible copper (Cu) wire inside it. The DBD reactor consists of a coaxial-cylindrical arrangement of a pin electrode covered by a closed end quartz tube placed inside a dielectric chamber. The diameter of the pin electrode is 3.95 mm. The quartz tube has inner and outer diameters equal to 4.0 mm and 6.0 mm, respectively. The length of the quartz tube that is inside the DBD reactor is 40 mm. A detailed description and the basic working principle of the plasma source can be found in \cite{kostov_transfer_2015, do_nascimento_different_2023}. In such configuration the plasma source operates with two discharge ignitions with the same gas passing through the system, being that the first one occurs inside the DBD chamber and the second one occurs at the distal end of the long tube, forming a plasma jet. In this work, the gas used to generate the plasmas was He (99.95\% pure, from Air Liquide, Brazil).

Measurements of discharge power were performed as a function of various parameters (gas flow rate ($Q$), length of the long tube ($L$), distance from outlet to target ($d$) and applied peak-to-peak voltage ($V_{pp}$)) for two different conditions: (i) without producing a plasma jet, that is, with the plasma jet off, and (ii) producing a plasma jet which impinges on a dielectric plate (plasma jet on). For the condition (i), the same Cu wire that passes through the long tube was extended and connected to the surface of the dielectric plate, as shown in the detail (i) of Fig.~\ref{expSetup}. In order to assure that a plasma jet would not be ignited in (i), $O_2$ gas (99.99\% pure, from Air Liquide, Brazi) was injected in the plastic connector at a flow rate of 0.5 slm. For the condition (ii), the only electrical connection between the plasma outlet and the dielectric plate is the plasma jet itself, as it is depicted in the detail (ii) of Fig.~\ref{expSetup}. In this case, there is no injection of $O_2$ in the plastic connector.

When using condition (i), since only the primary discharge is ignited, the measured power corresponds only to that dissipated in the plasma generated inside the reactor ($P_{Reactor}$). However, when using condition (ii), both the primary discharge and the plasma jet are ignited and, in this case, the power measurement returns the total power dissipated in both discharges ($P_{Total}$). Thus, under ideal conditions with occurrence of plasma discharges only in the reactor and in the jet, the power dissipated only by the plasma jet ($P_{Jet}$) can be estimated as the difference between $P_{Total}$ and $P_{Reactor}$, that is:

\begin{equation}
 {P_{Jet} = P_{Total} - P_{Reactor}}\label{eq08}
\end{equation}

However, eventually plasma discharges occur inside the plastic tube and, therefore, there is additional power dissipation along the tube. The configuration shown in Fig.~\ref{expSetup}, with the plastic connector for $O_2$ injection placed close to the position where the plasma jet is formed is intended to facilitate the separation of the power dissipated by the plasma jet from the power dissipated along the set formed by the reactor and the long tube. Even though, instead of estimating $P_{Jet}$, it is more appropriated to perform a more general analysis and express the difference between $P_{Total}$ and $P_{Reactor}$ in terms of a percentage of the total power:

\begin{equation}
 {Difference (\%) = \frac{P_{Total} - P_{Reactor}}{P_{Total}} \cdot 100\%}\label{eq09}
\end{equation}

\noindent
In this way, all power losses that are common for both discharges are implicitly disregarded in the data analysis.

The power supply employed to produce the plasmas was a commercial AC generator from GBS Elektronik GmbH (model Minipuls4) which works together with an arbitrary function generator from RIGOL (model DG1012) and a DC voltage generator. The DC supply was used to power the AC generator while the function generator was responsible for the voltage modulation. In this way, the set is able to produce a HV signal consisting of repetitive groups of sinusoidal waveforms (bursts), whose oscillation frequency ($f_{osc}$) was 29.0 kHz, followed by a voltage off interval. The entire HV signal repeated in a period $\tau$ = 1.7 ms.

Measurements of the applied voltage were performed using a 1000:1 voltage probe from Tektronix (model P6015A). The voltage across the 10 nF capacitor $C$ ($V_C (t)$) was measured using a 10:1 voltage probe from Tektronix (model P6139B). The waveform of the current that passes through the system was acquired using a current monitor from PearsonTM (model 4100). All the electrical signals were recorded simultaneously on a 200 MHz oscilloscope from Tektronix (model TDS2024B).

Optical emission spectroscopy (OES) measurements for the primary discharge were also performed in order to check if there were differences in its emission spectra when the plasma jet is off or on. For this purpose, it was employed a spectrometer from Avantes (model AvaSpec-ULS2048X64T), with spectral resolution (FWHM) equal to 0.76 nm, for measurements in the wavelength range from 200 nm to 750 nm. The light collection scheme, depicted in Fig.~\ref{expSetup}, is composed of a lens (25 mm in diameter) connected to an optical fiber.

\section{Results and discussion}\label{secResults}

\subsection{Method validation and proof of fundaments}\label{MetVal}

\begin{figure}
\centering
\includegraphics[width=0.75\linewidth]{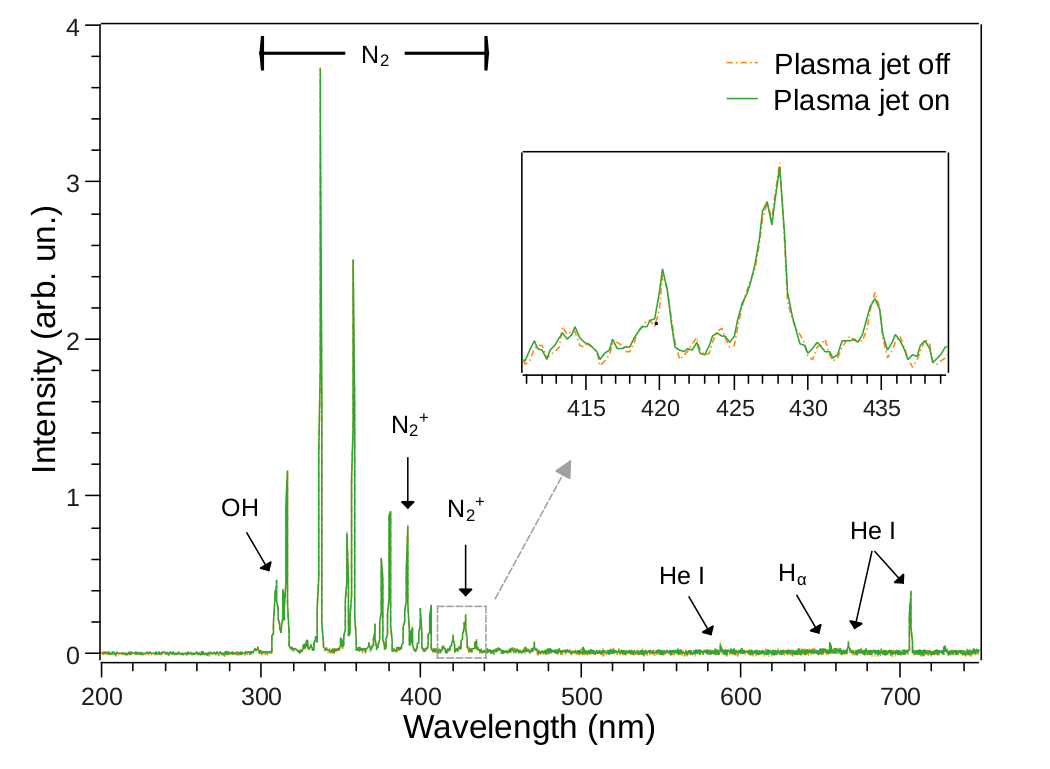}
\caption{Spectra measured for the light emission inside the reactor chamber with the plasma jet off (dashed line) and on (continuous line) for the wavelength range from 200 nm to 750 nm. The region from 411 nm and 439 nm, detached in the figure, is the one presenting the largest difference between the measured intensities.}
\label{reactorSpec}
\end{figure}

In order to check whether the ignition of the plasma jet affects or not the primary discharge, OES measurements were carried out in the plasma jet on and off conditions. The spectrum of the primary discharge with plasma jet on was measured first and, after that, the jet was turned off (by injecting the flow of $O_2$ in the tip of the long tube). Figure~\ref{reactorSpec} presents the measured spectra for both cases in the 200-750 nm wavelength range. As it can be seen, the two curves present no notable differences, even in the wavelength range from 411 nm to 439 nm, which is the most affected region by the plasma jet turning off. Such a result means that the existence of the plasma jet does not have a significant influence on the parameters of the primary discharge, since changes in any of the electrical parameters of the discharge in the reactor would probably lead to a change in the emission spectrum.

\begin{figure}
\centering
\includegraphics[width=\linewidth]{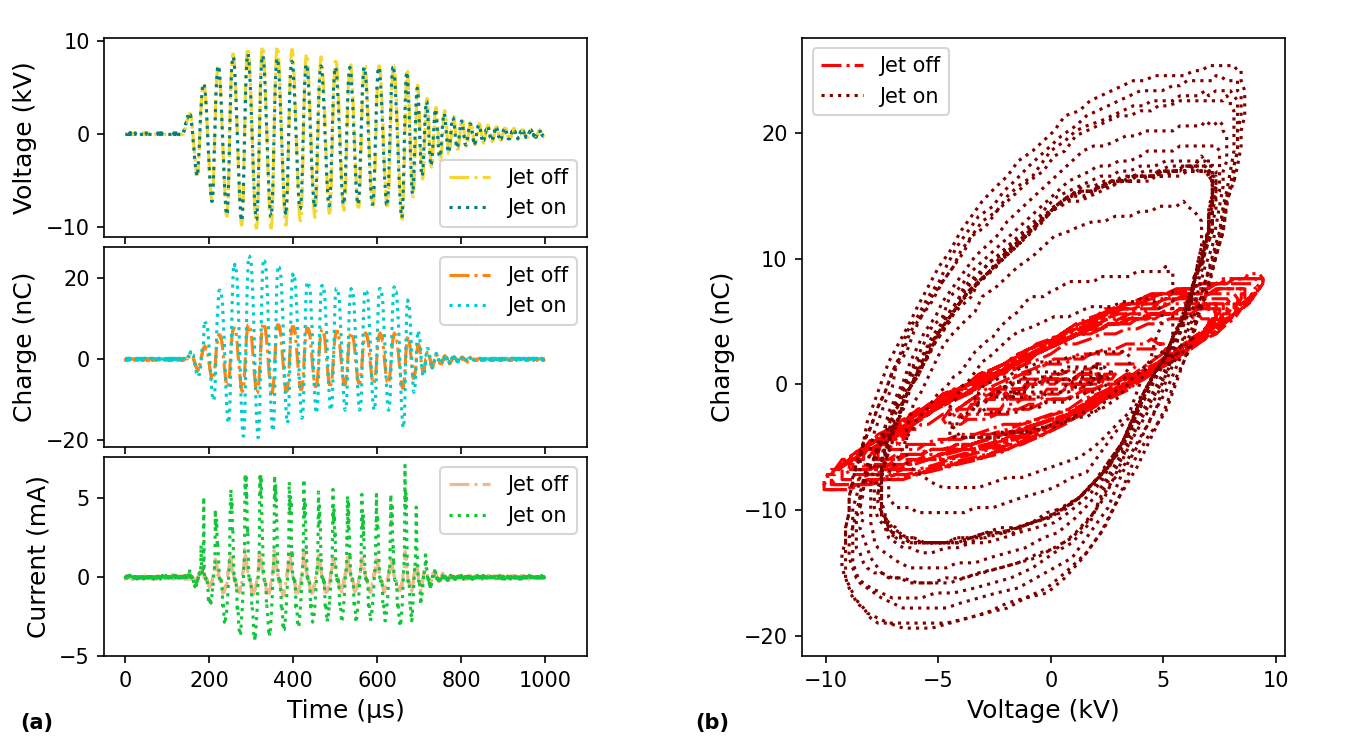}
\caption{Examples of (a) waveforms for voltage, charge and total current measured with the plasma jet on and off with (b) their corresponding $q-V$ plots. The operating conditions were $L$ = 100 cm, $Q$ = 2.0 slm and $d$ = 5.0 mm.}
\label{vqiWF}
\end{figure}

Examples of the waveforms of applied voltage, charge and total current under typical operating conditions ($L$ = 100 cm, $Q$ = 2.0 slm and $d$ = 5.0 mm) with the plasma jet off and on are shown in Fig.~\ref{vqiWF} (a). The associated $q-V$ plots are presented in~\ref{vqiWF} (b). By comparing the voltage waveforms with the plasma jet on and off, it can be seen that their temporal evolutions are very similar. However, when the plasma jet is ignited the $V(t)$ signal for the plasma on condition presents a small voltage drop, which is probably due to the higher discharge current, although the voltage drop is not proportional to the increase in current.

An important observation to be made about the loops in the $q-V$ curves shown in {Fig.~\ref{vqiWF}} (b) is that they are not stationary. That is, the area enclosed by each loop changes from cycle to cycle. This occurs due to the variation in the amplitude of $V(t)$, which is not constant within the burst, and the consequent variation in the $q(t)$ amplitude, as can be seen in {Fig.~\ref{vqiWF}} (a). Since the areas of all cycles are summed up when applying {equation~\ref{eq06}}, the variations in $q-V$ areas among the cycles within the burst do not affect the calculation of $P_{dis}$.

Regarding the waveforms for charge and current shown in Fig.~\ref{vqiWF}, it is noticeable that both $q(t)$ and $i(t)$ reach higher peak values when the plasma jet is on, nearly four times higher than in the plasma off case. The behavior of $q(t)$ presents slightly different shapes when comparing the plasma jet off and on cases. The resulting $q-V$ plots for the plasma off and on conditions are remarkably different in shape, being that the one with the plasma jet off slightly resembles a parallelogram, despite the rounded corners at negative and positive peak values, while the other in which the plasma jet is on approaches an ellipse. This reveals that the long tube configuration introduces resistive load components in the system, which partially confirms the transmission line model for the long tube proposed by Bastin \textit{et al} \cite{Bastin_optical_2020}.

A resistive load in an APPJ usually results from a corona discharge component in the plasma jet \cite{Kogelschatz_dielectric-barrier_2003}. The presence of metallic components inside the reactor chamber and the Cu wire inside the long tube are likely turning the primary discharge and plasma jet into a corona-like or corona-DBD discharge \cite{sakiyama_corona-glow_2006, kim_feasibility_2007}. Such behavior is clearly stronger in the plasma jet than in the primary discharge.

In addition to the difference in the shapes of the $q-V$ plots for the two conditions, it can also be seen that when the plasma jet is off, the figure tends to be symmetrical in the negative and positive voltage/charge phases. On the other hand, when the plasma jet is on, no symmetry is observed. In this case, it can be seen that the negative phase of the $q-V$ figure is more elliptical than the positive phase, which, in turn, means that the plasma jet in the negative phase is more resistive than in the positive one.

\begin{figure}
\centering
\includegraphics[width=0.75\linewidth]{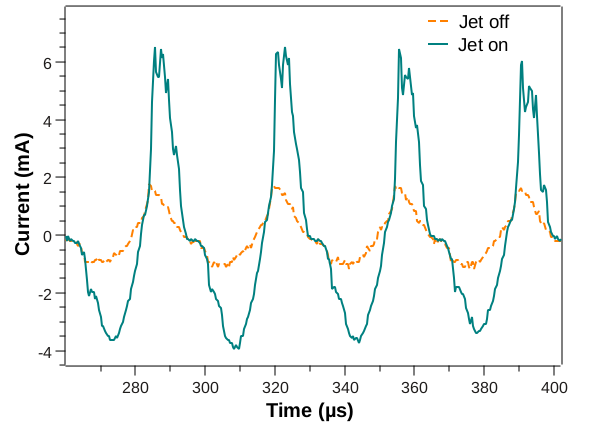}
\caption{Details of the measured currents with the plasma jet off (dashed line) and on (solid line).}
\label{iwfDetail}
\end{figure}

Regarding the measured current waveforms, some differences can also be seen when the plasma jet is off or on. In Fig.~\ref{iwfDetail} it can be seen that when the plasma jet is on the $i(t)$ presents more shaped positive peaks and, in addition, there are multiple spikes in each of the peaks. However, in the jet off condition there is only one current peak in each cycle, and it is rounded instead of shaped. Based on that, we can infer that each plasma discharge is running in a different mode, with the discharge inside the reactor being diffuse (characterized by the rounded peaks in $i(t)$) while the plasma jet being filamentary (because of the multiple spikes in each $i(t)$ sub-cycle).

\subsection{Discharge power and effective current measurements}\label{PowerCurrent}

\begin{figure}
\centering
\includegraphics[width=\linewidth]{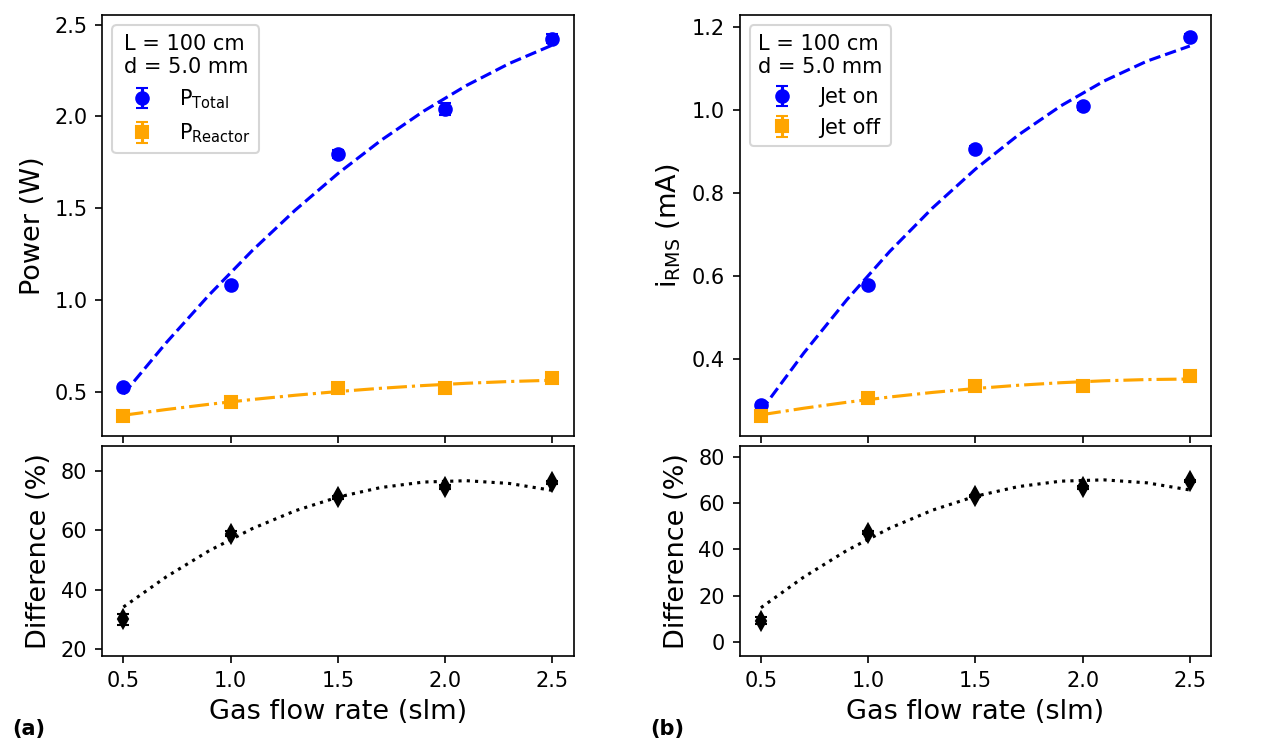}
\caption{Discharge parameters as a function of the gas flow rate. (a) Total power ($P_{Total}$) dissipated on the plasma source, with the plasma jet on, and the power dissipated only on the primary discharge ($P_{Reactor}$), with the jet off; the difference corresponds to the power dissipated on the plasma jet. (b) Effective currents measured with plasma jet on and off and the corresponding difference.}
\label{piVSgasflow}
\end{figure}

Figure~\ref{piVSgasflow} shows the variation of (a) discharge power and (b) effective currents measured as a function of the gas flow rate ($Q$). In that figure, $P_{Total}$ is the power dissipated in all parts of the plasma source, that is, with the plasma jet on, $P_{Reactor}$ is the power dissipated on the primary discharge plasma, with the plasma jet off.

The curves of effective currents ($i_{RMS}$) as a function of $Q$ are also shown for the conditions with the plasma jet on and off. The differences shown in the bottom of Fig.~\ref{piVSgasflow}(a and b) are presented for visualization of the behavior of the curves as a function of Q. Both power and effective current values were measured for the condition in which the length $L$ of the long tube was 100 cm and the distance $d$ between plasma outlet and target was 5.0 mm.

In the measured $Q$ interval, it was observed that the $P_{Total}$ values increased by a factor of five, almost linearly with $Q$, while the $P_{Reactor}$ ones increased less than 60\% in the same conditions. In this way, the difference increases significantly with the gas flow rate for $Q$ $\leq$ 1.0 slm, but its growth is slower for higher $Q$ values, showing a trend towards stabilization. The behavior of the $i_{RMS}$ values as a function of $Q$ are almost the same as the ones observed for the power values.

\begin{figure}
\centering
\includegraphics[width=\linewidth]{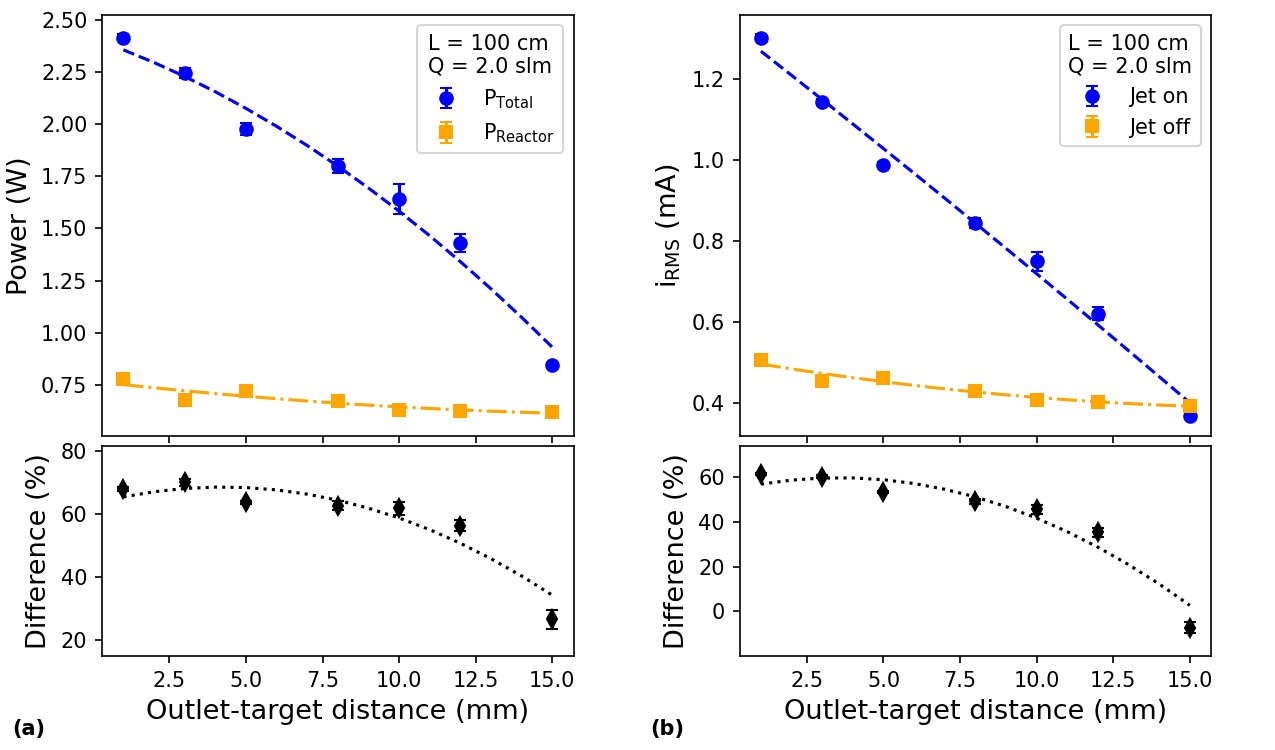}
\caption{Discharge parameters versus distance from outlet to target. (a) Power measurements and calculated difference. (b) Effective current measured and the calculated difference.}
\label{piVSdist}
\end{figure}

The behavior of the discharge parameters as a function of the distance $d$ between plasma outlet and target is presented in Fig.~\ref{piVSdist}. In (a) are shown the curves for $P_{Total}$, $P_{Reactor}$ and the difference as the percentile of $P_{Total}$. In (b) are shown the $i_{RMS}$ curves measured for the plasma on and off cases, and the corresponding difference for comparison of its behavior with the difference in power. As it can be seen from Fig.~\ref{piVSdist}(a), both $P_{Reactor}$ and $P_{Total}$ present a trend of reduction as $d$ is increased. However, the total reduction of $P_{Reactor}$ in the measured $d$ range is of the order of 20\%, while $P_{Total}$ decreases by a factor of ~2.8 in the same $d$ interval. It is important to note that the plasma jet does a transition from conducting to non-conducting mode (i.e. the plasma jet does not touch the target anymore) at a $d$ value between 12.0 and 15.0 mm, that is, in the measurement performed at $d$ = 15.0 mm the plasma jet is not in visible contact with the dielectric, behaving almost as a free jet. Such transition is more evident when looking at the curve for difference in the bottom of Fig.~\ref{piVSdist}(a), since the values present a smooth change up to $d$ = 12.0 mm, but a strong reduction occurs after that.

An important comment regarding the $P_{Reactor}$ measurements must be made here: the connection between the flexible Cu wire at the tip of the long tube and the dielectric plate has been made in a way that its length was kept the same for all $d$ values. This is important information which leads us to believe that the dielectric barrier due to the gas gap has a significant influence on the behavior of the primary discharge in the case where the plasma jet is off, even with the Cu wire connected to the dielectric plate.

\begin{figure}
\centering
\includegraphics[width=\linewidth]{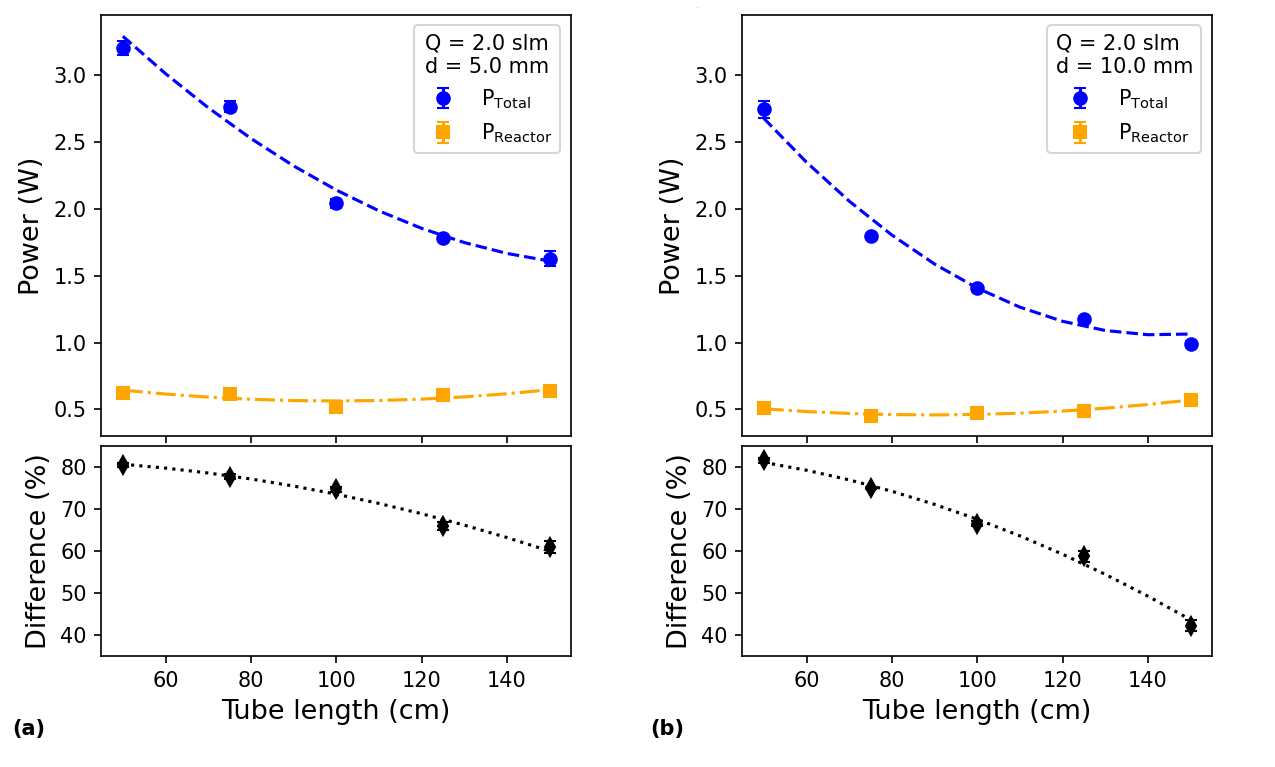}
\caption{Discharge power as a function of the length of the plastic tube. (a) Curves measured for $d$ = 5.0 mm. (b) Curves obtained for $d$ = 10.0 mm. $Q$ was 2.0 slm in both cases.}
\label{powerVStubelength}
\end{figure}

The influence of the length $L$ of the long tube on the power dissipation in different parts of the device was also evaluated. In Fig.~\ref{powerVStubelength} are presented the curves of $P_{Total}$, $P_{Reactor}$ and the corresponding difference between both curves as a function of $L$. Being that the measurements were performed for $d$ = 5.0 mm in (a) and for $d$ = 10.0 mm in (b). As it can be seen in Fig.~\ref{powerVStubelength}, the tube length does not significantly affect the $P_{Reactor}$ values. This is an expected result, since there are no changes in any parameters directly related to the primary discharge. On the other hand, the values of $P_{Total}$ as a function of the tube length present a monotonically decreasing trend as $L$ is increased, independently of the $d$ value. This happens because the potential at the tip of the Cu wire inside the long tube decreases as $L$ is increased \cite{kostov_transfer_2015, do_nascimento_comparison_2017}.

\begin{figure}
\centering
\includegraphics[width=\linewidth]{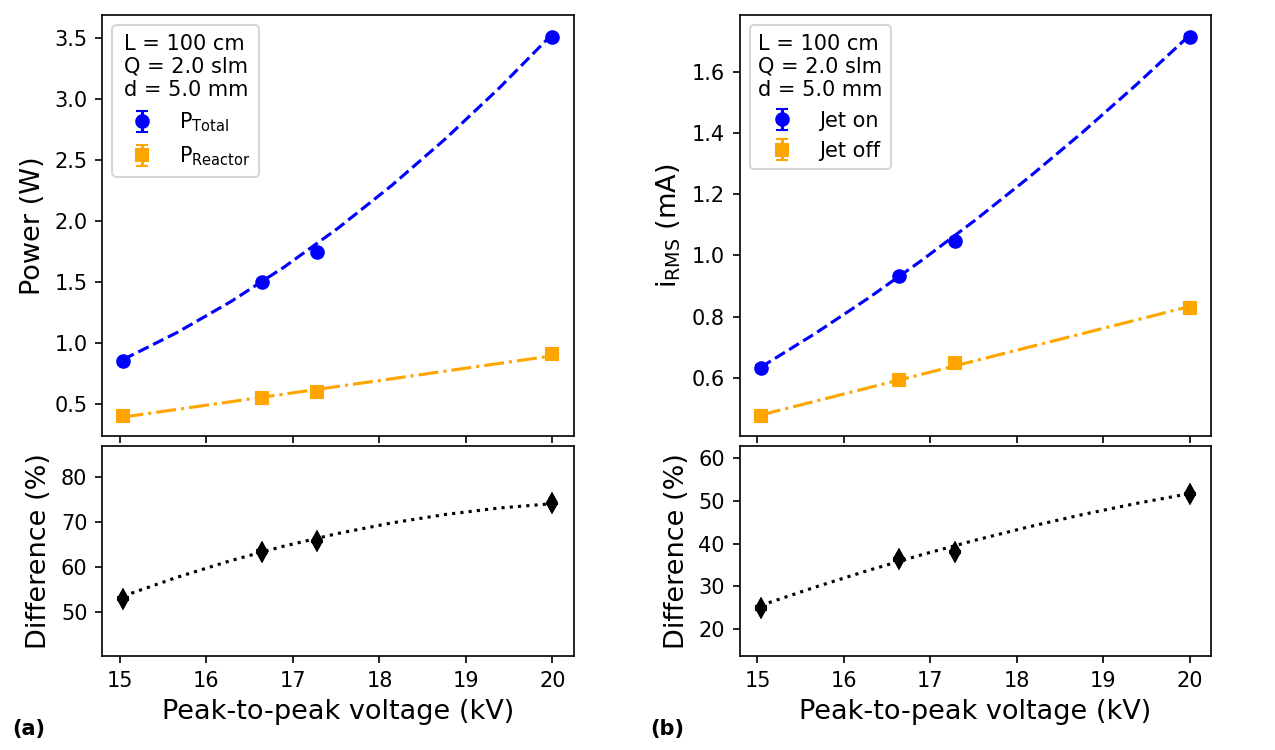}
\caption{Discharge parameters as a function of the applied peak-to-peak voltage. (a) Power measurements. (b) Effective currents.}
\label{piVSvoltage}
\end{figure}

Figure~\ref{piVSvoltage} presents measurements of discharge power and effective current as a function of the peak-to-peak voltage value ($V_{pp}$). As expected, the discharge power and effective current values increase when $V_{pp}$ is increased, in both experimental arrangements. However, it is noticeable that the difference between $P_{Total}$ and $P_{Reactor}$ increases considerably as $V_{pp}$ is incremented, and the same happens with the $i_{RMS}$ values obtained for the conditions with the plasma jet off and on.

From the $i_{RMS}$ vs $V_{pp}$ curves in Fig.~\ref{piVSvoltage} (b) it can be seen that when the plasma jet is off, the inclination of the curve is constant, from which can be inferred a constant impedance of the system in such a condition. On the other hand, when the plasma jet is on, the behavior of the $i_{RMS}$ vs $V_{pp}$ curve appears to be quadratic, with an electrical admittance that increases with $V_{pp}$, which implies a reduction in $Z$ as the applied voltage is increased. Such a reduction in $Z$ with $V_{pp}$ being increased is compatible with a better ignition of the plasma jet, which causes a reduction in its resistivity.

\subsection{q-V plot analysis and capacitance estimations}\label{qVplotAn}
This section is intended to provide experimental data about the influence of each parameter in the $q-V$ plot shapes as well as in the capacitance values for various working conditions. These data presented here can potentially help in the understanding of the electrical equivalent circuit for different setups and working conditions.

\subsubsection{Influence of the gas flow rate}\label{gfrInflu}

\begin{figure}
\centering
\includegraphics[width=\linewidth]{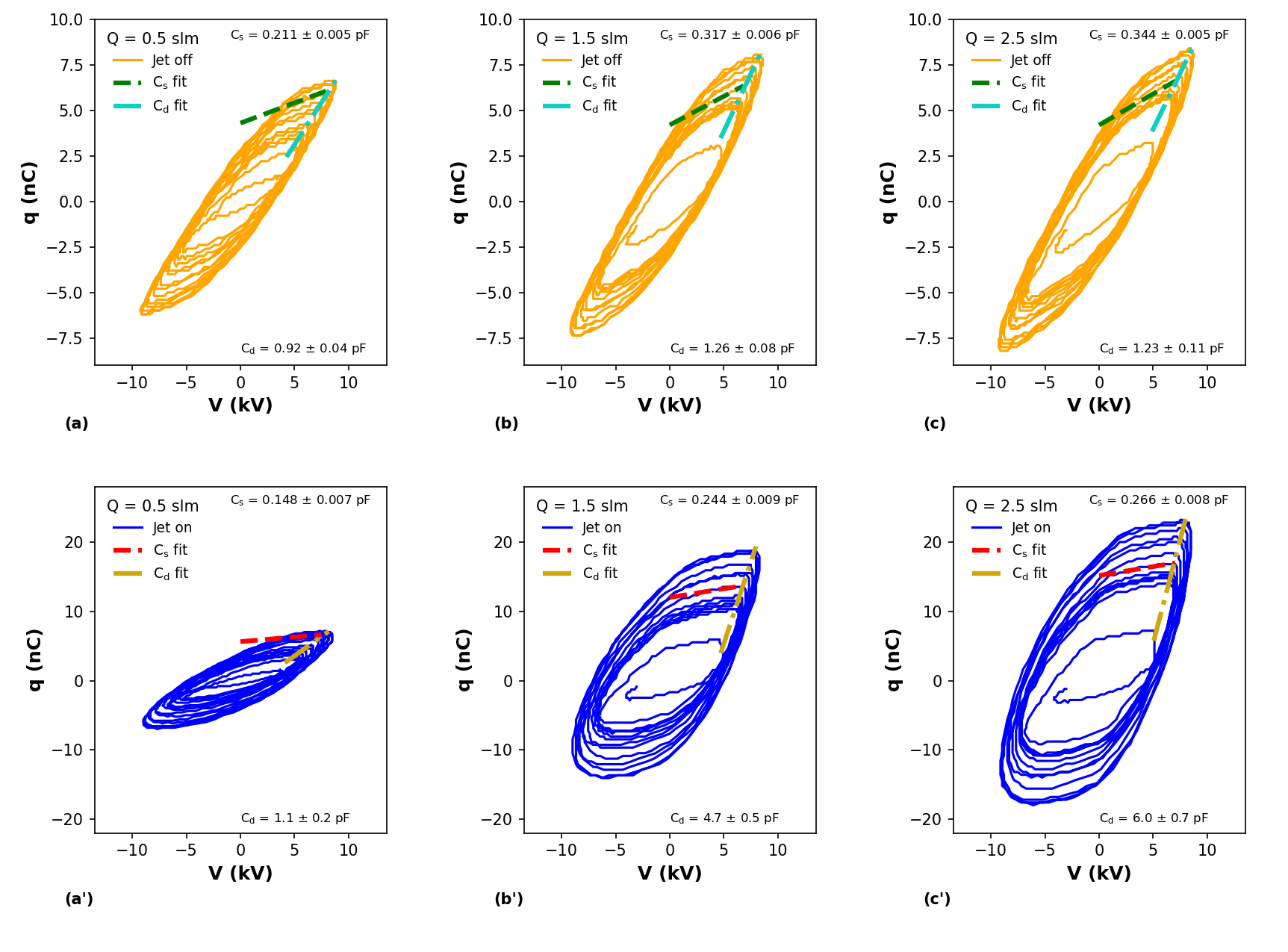}
\caption{$q-V$ plots obtained for different $Q$ values (indicated on the graphs) with the plasma jet off (a, b, c) and on (a’, b’, c’). $L$ = 100 cm and $d$ = 5.0 mm for these data.}
\label{qVvsFlow}
\end{figure}

Figure~\ref{qVvsFlow} shows the $q-V$ plots obtained with the plasma jet off (a, b, c) and on (a’, b’, c’) for $L$ = 100 cm, $d$ = 5.0 mm and different $Q$ values (indicated on the graphs). For each $q-V$ cycle, $C_s$ and $C_d$ values were estimated from the $q-V$ plots. Ten cycles, in the middle of the discharge, were used to calculate the average values of $C_s$ for each working condition. The $C_d$ values were obtained by fitting the data formed by the ($V_{q_{max}}$, $q_{max}$) curve in 10 cycles, similarly to the method described by Pipa \textit{et al} \cite{Pipa_role_2013}. An example of the fitted $C_s$ and $C_d$ curves are presented in each $q-V$ plot in Fig.~\ref{qVvsFlow}, as well as the average values of $C_s$ and the $C_d$ obtained in each case.

As it can be seen in Fig.~\ref{qVvsFlow} (a-c), when the plasma jet is off, the shape of the $q-V$ plots does not change significantly for different $Q$ values and, in all cases, it resembles a parallelogram. However, as $Q$ increases, it can be noticed that the corner of minimum V and q values becomes more rounded, indicating a trend of a more resistive discharge in this phase. Regarding the $q-V$ plots when the plasma jet is on (a’- c’), the shapes change significantly when $Q$ is increased from 0.5 slm to 1.5 slm, from a quite elliptical form to mixed one (elliptical for negative q values and close to a parallelogram for positive $q$). A trend for a transition from the mixed shape to a parallelogram can be observed when $Q$ is increased from 1.5 slm to 2.5 slm, from which can be inferred that an increase in the gas flow rate makes the discharge less resistive. In fact, for large $Q$ values the shape of the $q-V$ plot with the plasma jet on resembles an hexagon.

\begin{figure}
\centering
\begin{minipage}[c]{0.46\textwidth}
\centering
\includegraphics[width=\textwidth]{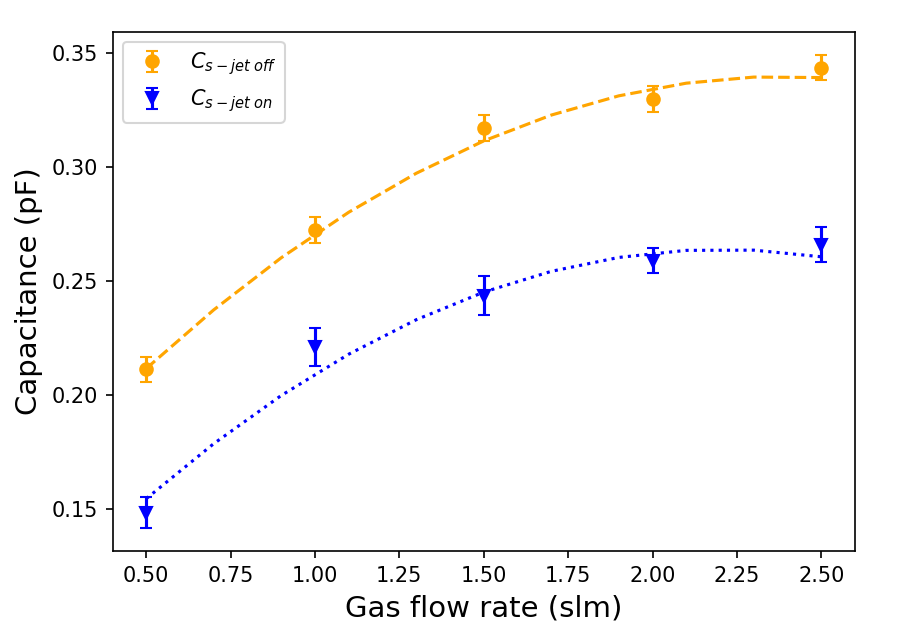}
\caption{Variation of $C_s$ values as a function of the gas flow rate.}\label{CsvsFlow}
\end{minipage}
\hfill
\begin{minipage}[c]{0.46\textwidth}
\centering
\includegraphics[width=\textwidth]{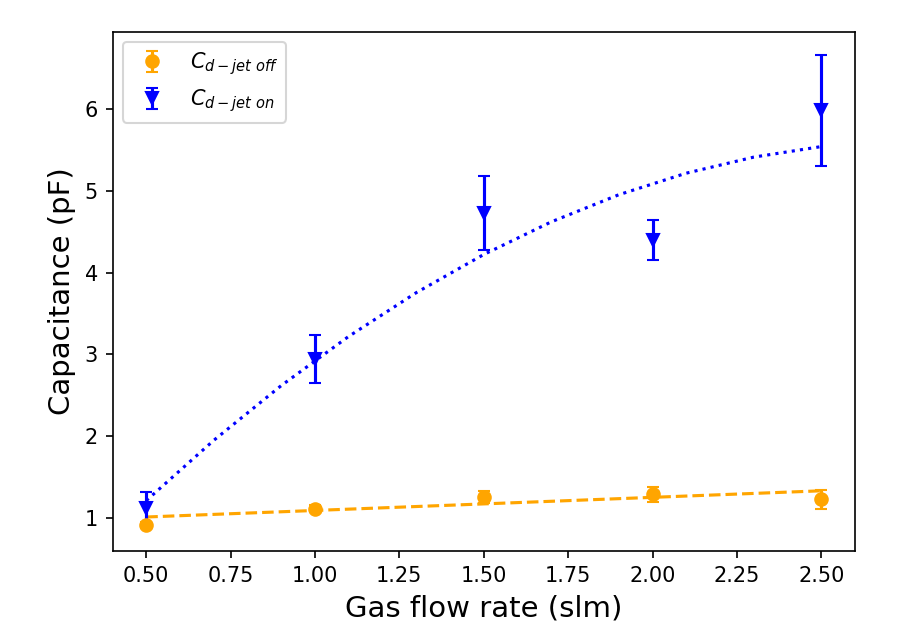}
\caption{Variation of $C_d$ values as a function of the gas flow rate.}\label{CdvsFlow}
\end{minipage}
\end{figure}

The variation of $C_s$ and $C_d$ values as a function of $Q$ is shown in Figs.~\ref{CsvsFlow} and~\ref{CdvsFlow}, from which can be seen that the capacitance values have a tendency to increase as a function of $Q$ in all cases, with the $C_d$ values for plasma plume off presenting only a small variation. The increase of the $C_s$ values as a function of $Q$ may be related to an increase of the capacitance value of the gas gap ($C_g$), since at higher $Q$ the density of He atoms in the gas gap region increases, leading to an increment in the relative permittivity of the He gas.

The $C_d$ curves as a function of $Q$ shown in Fig.~\ref{CdvsFlow} present different variation rates in the cases where the plasma jet is off or on. This is due to the fact that with plasma jet off the obtained $C_d$ values are only related to the dielectric that covers the electrode in the reactor and this is the reason for the almost constant $C_d$ values as a function of $Q$. When the plasma jet is on, $C_d$ values of the reactor are summed up to the capacitances distributed along the long tube, which are probably arranged in parallel.

\subsubsection{Influence of the outlet-to-target distance}\label{odInflu}

\begin{figure}
  \centering
  \includegraphics[width=\linewidth]{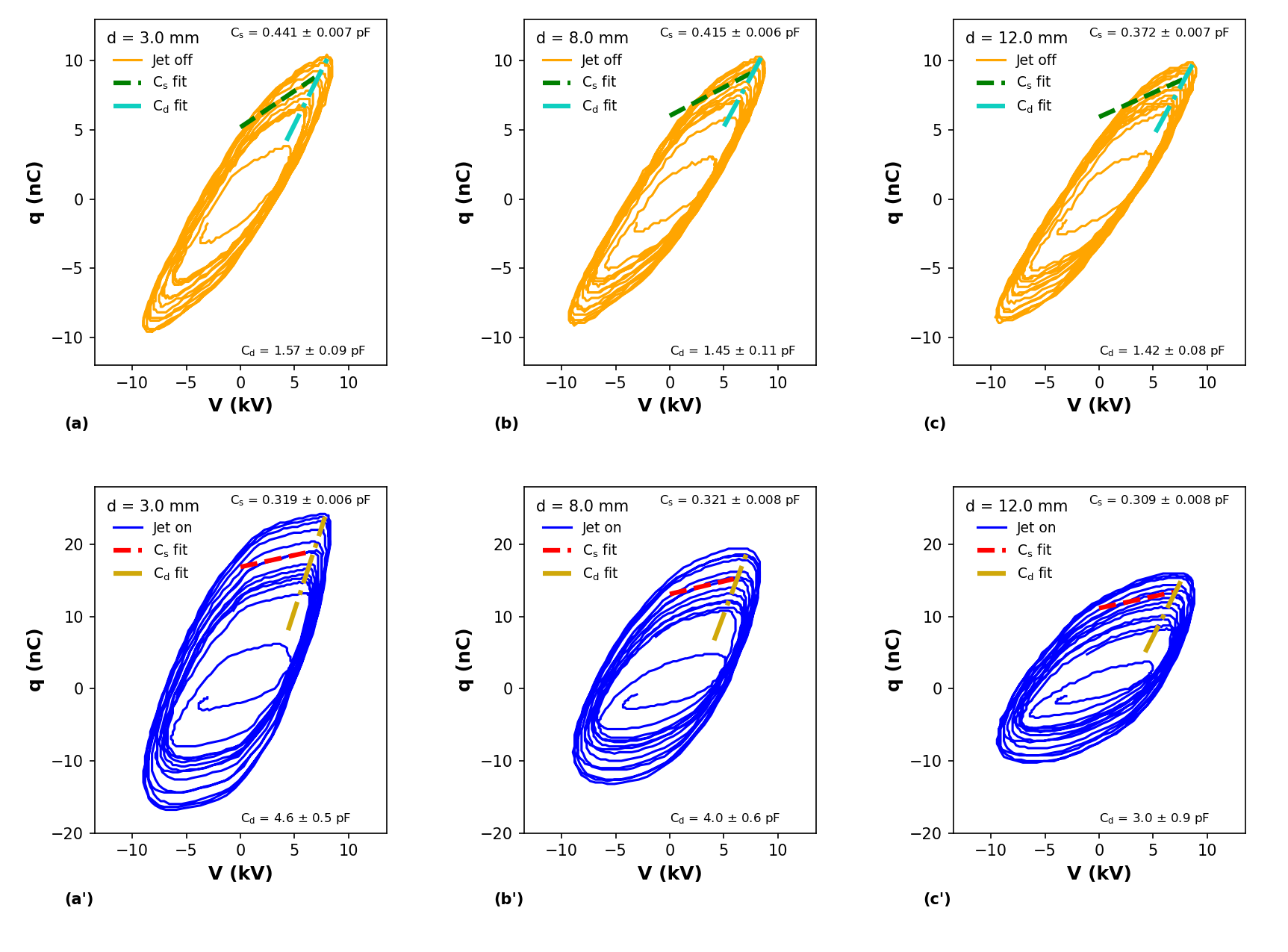}
  \caption{$q-V$ plots obtained for different $d$ values (indicated on the graphs) with the plasma jet off (a, b, c) and on (a’, b’, c’). $L$ = 100 cm and $Q$ = 2.0 slm for these data}
  \label{qVvsDist}
\end{figure}

Figure~\ref{qVvsDist} shows $q-V$ plots obtained for different $d$ values with the plasma jet off (a-c) and on (a’-c’). Variations in the $q-V$ shape are observed in both cases when $d$ is modified. When the plasma jet is off, it can be seen that the $q-V$ curves are more elliptical for small $d$ values and more parallelogram-like for large d. However, when the plasma jet is on, the opposite occurs, that is, the $q-V$ diagram has hexagonal features for small $Q$ and elliptical shape for large $Q$. The last is expected, since the conductivity of the plasma jet decreases when $d$ is increased, leading to a more resistive discharge.

\begin{figure}
\centering
\begin{minipage}[c]{0.46\textwidth}
\centering
\includegraphics[width=\textwidth]{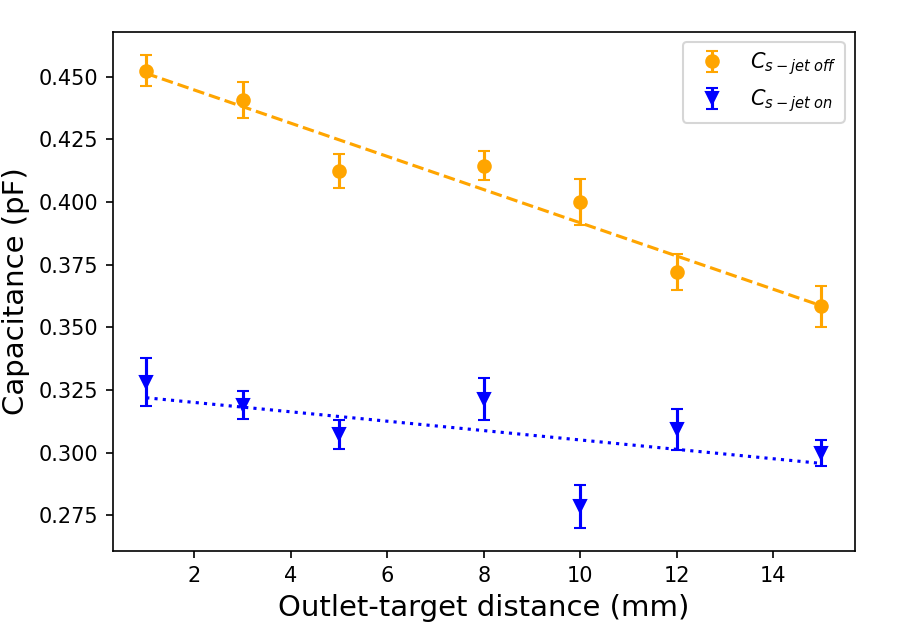}
\caption{Variation of $C_s$ values as a function of the distance between plasma outlet and target.}\label{CsvsDist}
\end{minipage}
\hfill
\begin{minipage}[c]{0.46\textwidth}
\centering
\includegraphics[width=\textwidth]{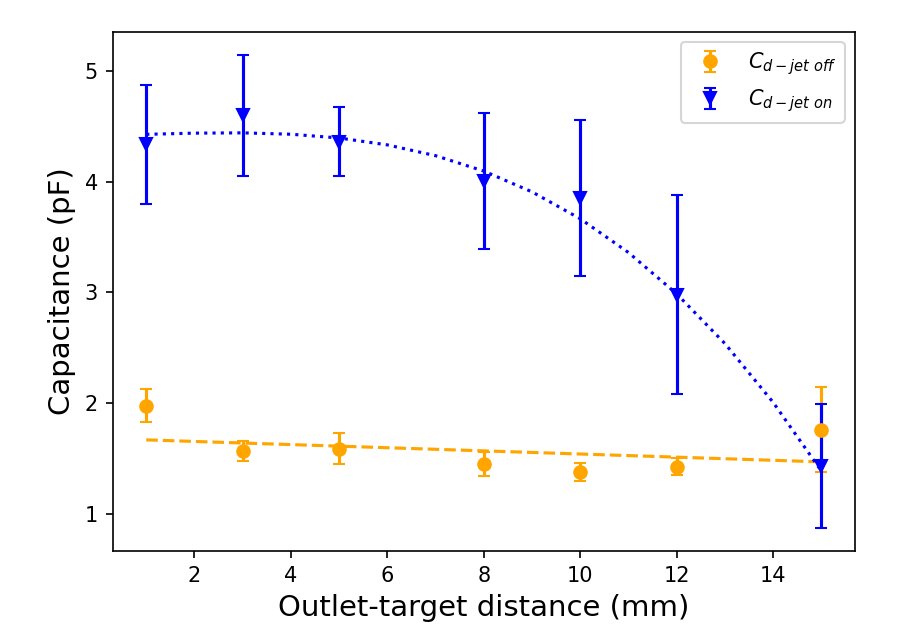}
\caption{Variation of $C_d$ values as a function of the distance between plasma outlet and target.}\label{CdvsDist}
\end{minipage}
\end{figure}

The curves of the $C_s$ and $C_d$ values as a function of $d$ are presented in Figs.~\ref{CsvsDist} and~\ref{CdvsDist}, respectively. From Fig.~\ref{CsvsDist} it can be seen that the $C_s$ values tend to decrease as $d$ is increased. Such a reduction trend is consistent with the decrease of $C_g$ value as $d$ is increased. The $C_s$ values obtained with the plasma jet on also present a downward trend as $d$ is increased.

In Fig.~\ref{CdvsDist} it can be seen that the $C_d$ values do not change significantly with $d$ for the jet off condition. In fact, the $C_d$ values in this case are consistent with the ones measured as a function of $Q$, that is, the $C_d$ values obtained with the plasma jet off lies in the 1-2 pF range. In the condition with the plasma jet on, the $C_d$ values present a small decrease trend for $d$ $>$ 10 mm and a more pronounced reduction after that. This happens because at large $d$ values the plasma jet enters in a transition to non-conducting mode and the $C_d$ values tend to be the same as the one obtained with the plasma jet off.

\subsubsection{Influence of tube length}\label{tlInflu}

\begin{figure}
  \centering
  \includegraphics[width=\linewidth]{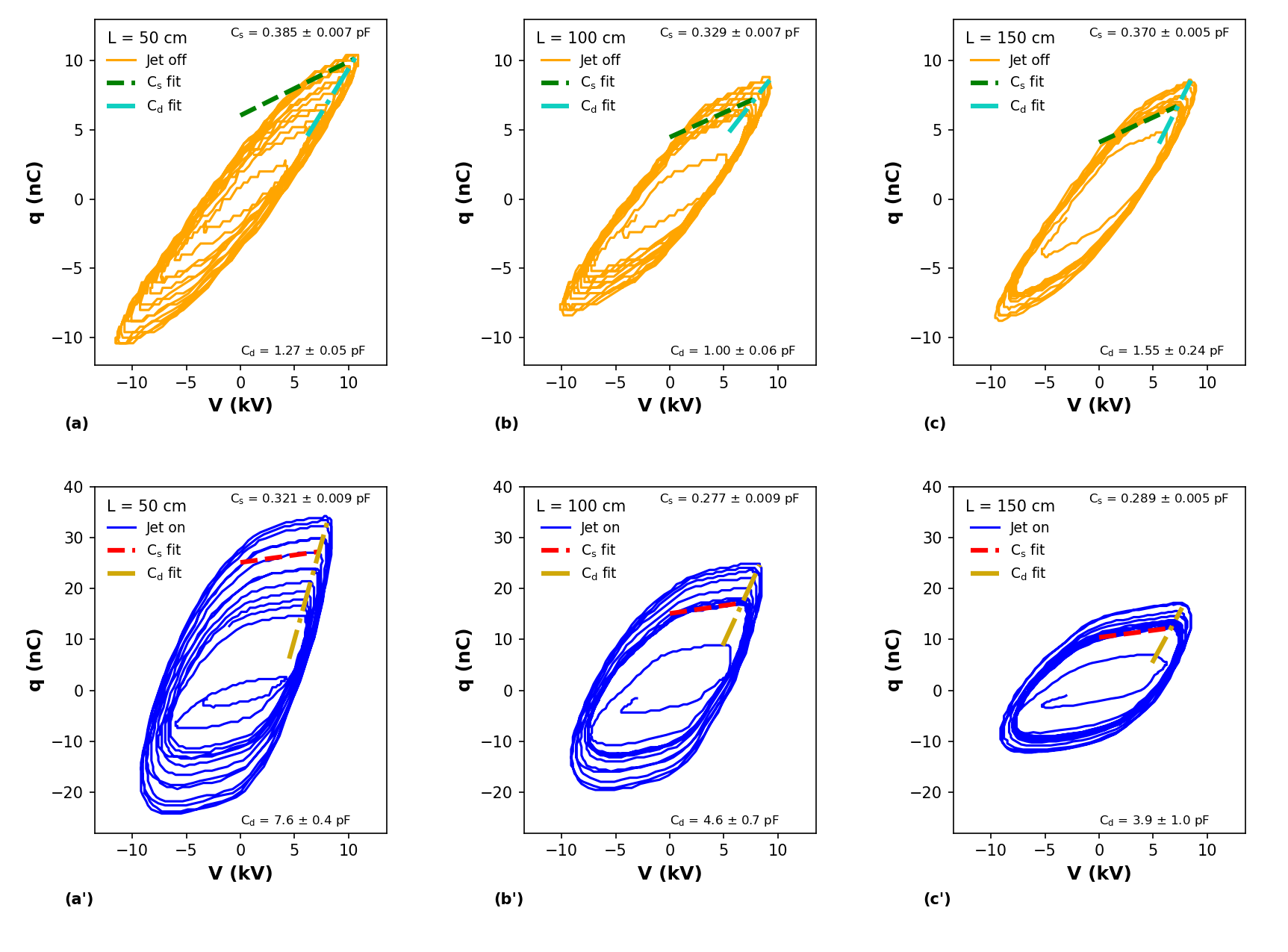}
  \caption{$q-V$ plots obtained for different tube lengths (indicated on the graphs) with the plasma jet off (a, b, c) and on (a’, b’, c’). $Q$ = 2.0 slm and $d$ = 5.0 mm for these data.}
  \label{qVvsTLength}
\end{figure}

The $q-V$ plots obtained for different $L$ values with the plasma jet off and on are shown in Fig.~\ref{qVvsTLength}, from which can be seen that the variation of the tube length does not cause a significant change in the shapes of the $q-V$ figures. However, it is noticeable that the tube length affects the amount of charge stored in each cycle, that is, for large $L$ values, the $q-V$ curves tend to overlap in different cycles while for small $L$ the $q-V$ curves present a wider range of amplitudes. This effect is more pronounced in the jet off condition.

\begin{figure}
\centering
\begin{minipage}[c]{0.46\textwidth}
\centering
\includegraphics[width=\textwidth]{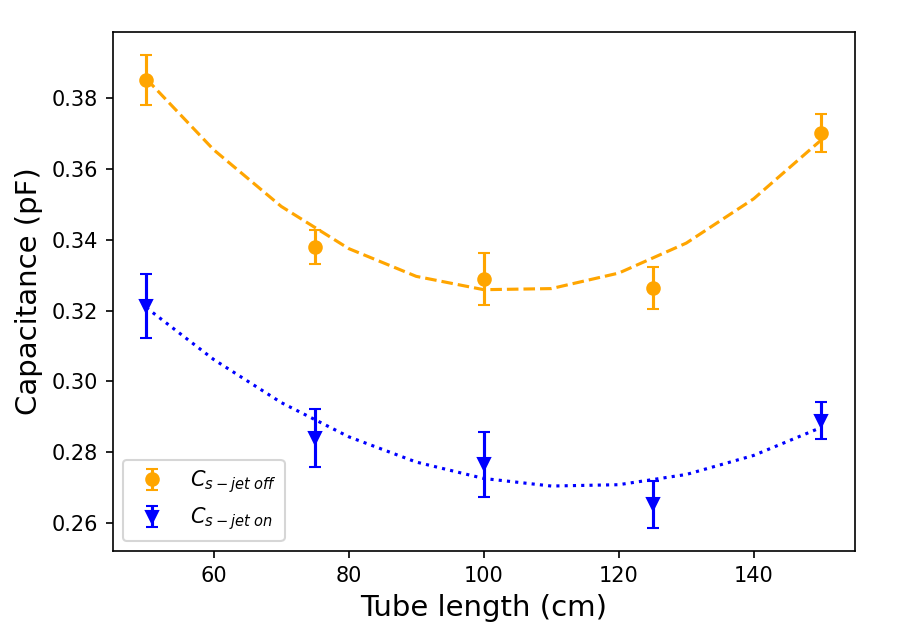}
\caption{Variation of $C_s$ values as a function of the tube length}\label{CsvsTLength}
\end{minipage}
\hfill
\begin{minipage}[c]{0.46\textwidth}
\centering
\includegraphics[width=\textwidth]{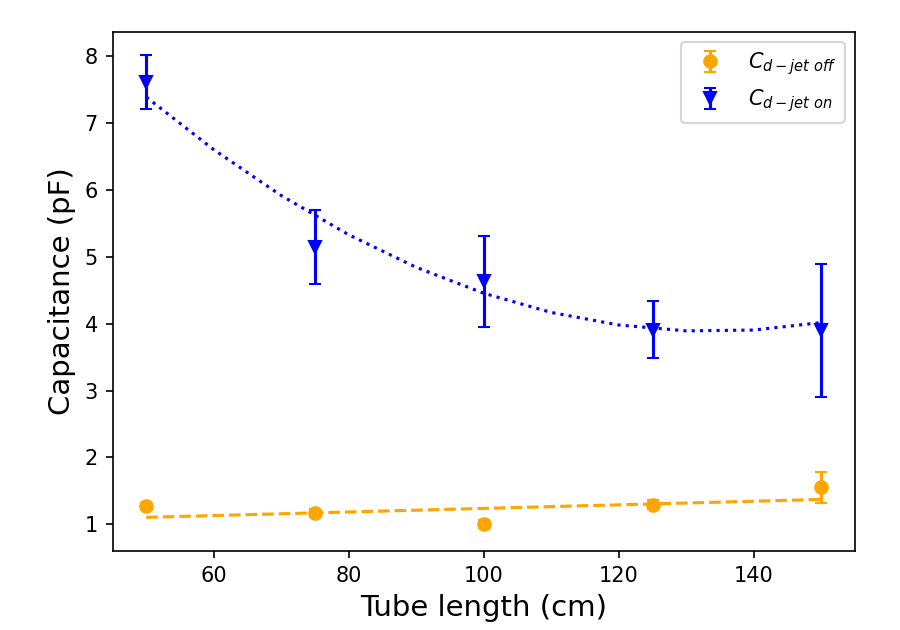}
\caption{Variation of $C_d$ values as a function of the tube length}\label{CdvsTLength}
\end{minipage}
\end{figure}

The behavior of $C_s$ and $C_d$ values as a function of $L$ are shown in Figs.~\ref{CsvsTLength} and~\ref{CdvsTLength}, respectively. An interesting observation about the $C_s$ values as a function of $L$ is that apparently there is an $L$ value that minimizes $C_s$ when keeping all the other parameters unchanged. However, the variation in $C_s$ values as a function of $L$ is not that large (~10-15\% for the given $L$ range), thus, the contribution of a reduction in this parameter to the electrical equivalent circuit may not be so significant.

In relation to the $C_d$ values as a function of $L$, it can be seen that the variation of this parameter with the plasma jet off is very small so it can be assumed that it remains unchanged as expected, since, in this case, the $C_d$ value refers to the dielectric inside the reactor only. On the other hand, when the plasma plume is on the $C_d$ values decrease as $L$ is increased.

\subsubsection{Influence of peak-to-peak voltage}\label{vppInflu}

\begin{figure}
  \centering
  \includegraphics[width=\linewidth]{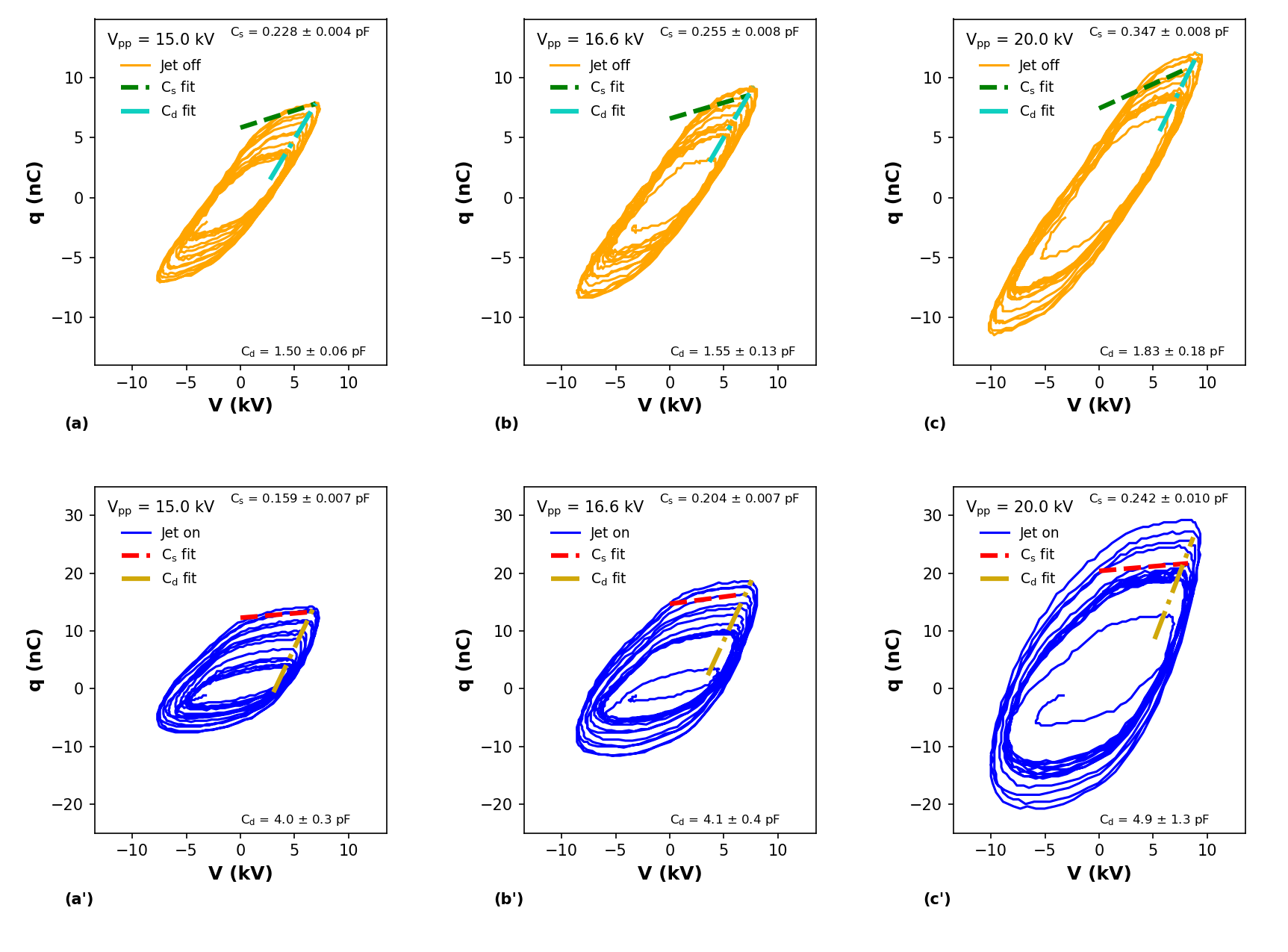}
  \caption{$q-V$ plots obtained for different peak-to-peak voltage values (indicated on the graphs) with the plasma jet off (a, b, c) and on (a’, b’, c’). $L$ = 100 cm, $Q$ = 2.0 slm and $d$ = 5.0 mm for these data.}
  \label{qVvsVpeak}
\end{figure}

The $q-V$ curves measured for different $V_{pp}$ values are shown in Fig.~\ref{qVvsVpeak} with the plasma jet off (a-c) and on (a’-c’). It can be seen that the applied voltage does not significantly change the shape of the $q-V$ curves when the plasma jet is off. Only an overlapping of the curves in different cycles is noted when $V_{pp}$ is equal to 20 kV. Such curve overlapping also happens at $V_{pp}$ = 20 kV when the plasma jet is on. Furthermore, at this working voltage with the plasma jet on a change in the shape of the $q-V$ plot is evident, being more elliptical than when working at lower $V_{pp}$ values.

\begin{figure}[htb]
\centering
\begin{minipage}[c]{0.46\textwidth}
\centering
\includegraphics[width=\textwidth]{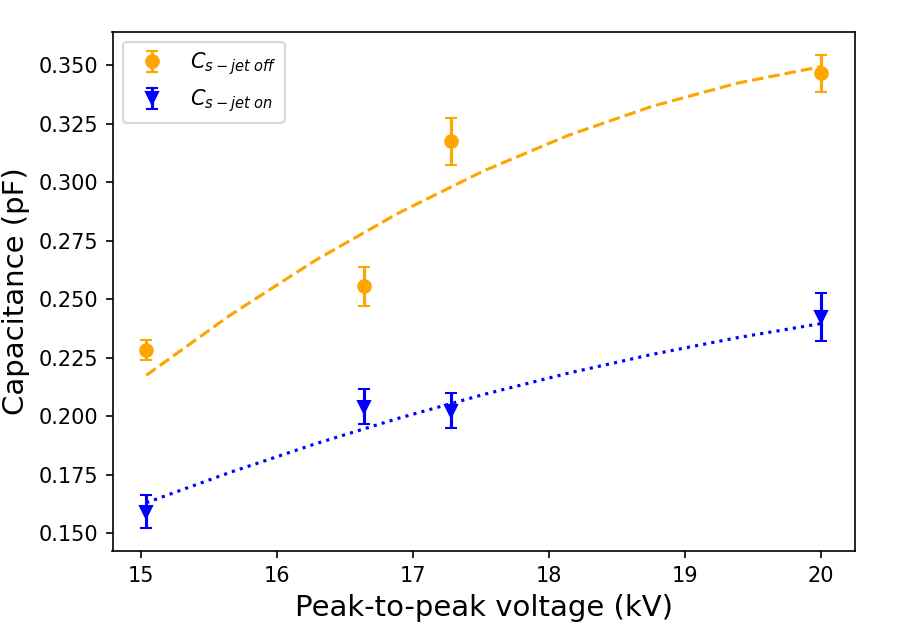}
\caption{Variation of $C_s$ values as a function of the peak-to-peak voltage.}\label{CsvsVpeak}
\end{minipage}
\hfill
\begin{minipage}[c]{0.46\textwidth}
\centering
\includegraphics[width=\textwidth]{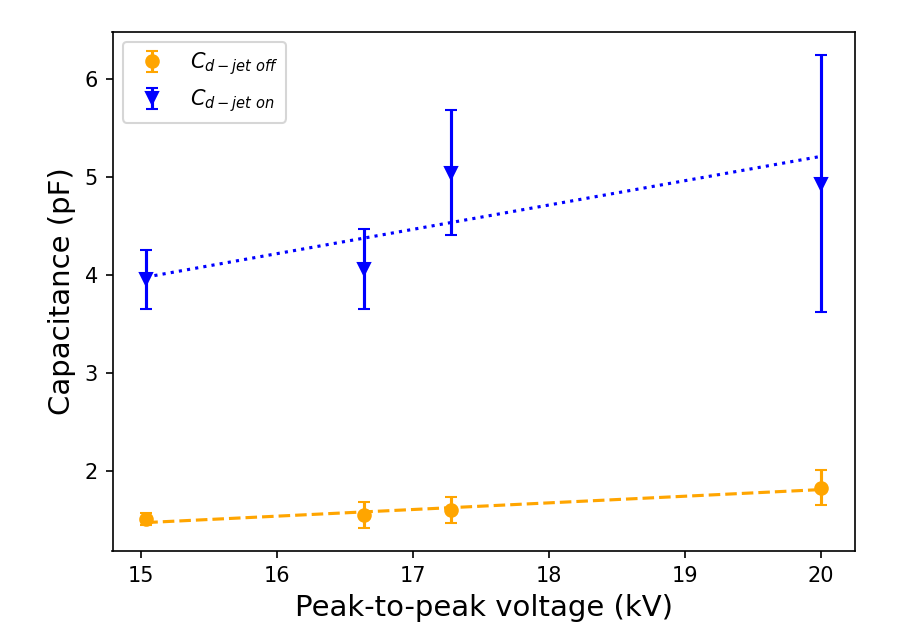}
\caption{Variation of $C_d$ values as a function of the peak-to-peak voltage.}\label{CdvsVpeak}
\end{minipage}
\end{figure}

The variation of $C_s$ and $C_d$ values as a function of $V_{pp}$ are shown in Figs.~\ref{CsvsVpeak} and~\ref{CdvsVpeak}, respectively. The $C_s$ values present a growth trend in the measured voltage range for both plasma off and on conditions. This result is probably related to an increase in the plasma volume inside the reactor and also with ignition of plasma discharges inside the long tube along its length, which are consequences of the higher values of the applied voltage. Regarding the $C_d$ values, they present only a very small growth trend as $V_{pp}$ is increased, in both plasma off and on cases, and can be considered constant.

\subsubsection{Comments on $q-V$ plot shapes and measured capacitance values}\label{qvComm}

The $q-V$ plot analysis revealed that the resistive load of the discharges is much higher at negative voltage polarities. However, this effect is not evident when only the primary discharge is ignited. That is, the (–,–) quadrant of the $q-V$ plots presented a very rounded shape when both primary discharge and plasma jet were ignited, but such shape was not that rounded without it. On the other hand, the (+,+) quadrant of the $q-V$ discharge tends to resemble a corner of a parallelogram, even with the ignition of the plasma jet. This is an indication that the electrical equivalent circuit model of the system under study should be split in two parts, one for each phase of the $q-V$ plot.

In general, all the measured capacitance values, especially $C_s$, presented some variation when the parameter under study was changed. The most significant variations in both $C_s$ and $C_d$ values were observed when changing the gas flow rate, followed by the outlet-to-target distance. $C_s$ also presented a significant variation as a function of the applied voltage.

The observed variations in $C_s$ values under different system configurations can be explained in different ways. However, in most cases this is related to a change in the electrical equivalent circuit model provided by the setup assembly and operating conditions.

Regarding the $C_d$ measurements with the plasma jet off, the obtained values are approximately constant for all the parameters under study in this work, with results lying between 1 and 2 pF. The portion of the quartz tube inside the DBD reactor shown in Fig.~\ref{expSetup} would provide a nominal capacitance value of ~25 pF if its inner and outer surfaces were entirely covered with a conducting material. Thus, since the electrode inside the tube has physical contact with the inner surface of the tube (the electrode’s diameter fits well the tube’s wall), the differences in the measured capacitance values come from the low effective surface contact between the plasma and tube’s outer wall, from which can be inferred that the measured $C_d$ values with the plasma jet off are the effective ones, not the nominal values. Therefore, $C_d$ can change according to the working condition.

\section{Conclusion}\label{secConclusions}

In this work, it was investigated power dissipated in a device that ignites two plasma discharges, one inside a DBD reactor and other as a plasma jet. This was done in a way that it was possible to know the power dissipated in the DBD reactor and the total power dissipated by the whole system. All the measurements were performed as a function of different parameters that could possibly interfere in the discharge parameters. It was verified that the power dissipated in the DBD reactor remains approximately unchanged when varying the gas flow rate, the length of the long tube and the distance between outlet and target. On the other hand, the power dissipated in the plasma jet changed significantly as a function of those parameters, increasing as a function of the gas flow rate, and decreasing as a function of the other two. It is interesting to note that the difference between the total dissipated power and the power dissipated in the reactor is higher than 50\% in most cases. This is an indication that the power dissipated by the plasma jet is much higher than in the reactor.

The electrical characteristics of the plasma source were also analyzed in this work. Based on the measured $q-V$ plot shapes, it was verified that the plasma jet discharges have a highly resistive component in the negative voltage polarities, which produce a very rounded corner in the $q-V$ plots.

Both system and dielectric capacitance ($C_s$ and $C_d$) values were estimated from the $q-V$ diagrams, with and without ignition of the plasma jet. Those measurements were performed as a function of the same parameters that were varied in power measurements. It was found that the $C_s$ values change in almost all cases, while significant changes in $C_d$ were observed only in the plasma jet on condition. In the situations where the $C_d$ values presented variations, the most likely explanation for the change is a stronger ignition of the plasma discharges, either inside the reactor or in the long tube, which leads to more effective contact of the plasma with the walls of the dielectrics involved, which in turn leads to higher effective $C_d$ values.

As a general conclusion, we can say that the EEC models proposed in the literature so far do not separately describe the system presented in this work. It is necessary to associate models reported in different works for each part of the discharge, specially when the plasma jet is ignited. This opens perspectives for future studies that allow a better understanding of the system as a whole.

\backmatter

\bmhead{Acknowledgements}

This work was supported by the São Paulo Research Foundation-FAPESP (grants 2019/05856-7 and 2020/09481-5) and by the Coordination for the Improvement of Higher Education Personnel–CAPES.

\bmhead{Author contributions}
The experiment was proposed by F.d.N., T.M.C.N. and K.G.K., and F.d.N. and K.A.P. executed the experiment. Data analysis was performed by F.d.N. and K.A.P. Original draft was written by F.d.N., T.M.C.N., K.A.P and K.G.K. All the authors contributed with discussions as well to the final version of the manuscript.

\noindent\textbf{Funding} This work received financial support from the São Paulo Research Foundation – FAPESP and from the Coordination for the Improvement of Higher Education Personnel – CAPES.

\noindent\textbf{Conflict of interest/Competing interests} The authors declare no Conflict of interest or Competing interests.



\noindent\textbf{Data availability} The data are contained in this manuscript. Raw data are available from the authors under reasonable request.



\bibliography{Discharge_power_dissipated_in_devices_with_double_ignition}

\end{document}